| | |
|---|---|
| Title | **The influence of power and frequency on the filamentary behavior of a flowing DBD—application to the splitting of $CO_2$** |
| Authors | A Ozkan[1,2], T Dufour[1], T Silva[3], N Britun[3], R Snyders[3], A Bogaerts[2] and F Reniers[1] |
| Affiliations | [1] Université Libre de Bruxelles, Chimie analytique et chimie des interfaces (CHANI), Campus de la Plaine, Bâtiment A, CP255, boulevard du Triomphe, 1050 Bruxelles, Belgium<br>[2] Universiteit Antwerpen, Research group PLASMANT, Universiteitsplein 1, 2610 Antwerpen-Wilrijk, Belgium<br>[3] Université de Mons, Chimie des Interactions Plasma-Surface (ChIPS), CIRMAP, 23 Place du Parc, 7000 Mons, Belgium, Materia Nova Research Center, Parc Initialis, B-7000 Mons, Belgium |
| Ref. | Plasma Sources Science & Technology, 2016, Vol. 25, No 2, 025013 (11pp) |
| DOI | http://dx.doi.org/10.1088/0963-0252/25/2/025013 |
| Abstract | In this experimental study, a flowing dielectric barrier discharge operating at atmospheric pressure is used for the splitting of $CO_2$ into $O_2$ and CO. The influence of the applied frequency and plasma power on the microdischarge properties is investigated to understand their role on the $CO_2$ conversion. Electrical measurements are carried out to explain the conversion trends and to characterize the microdischarges through their number, their lifetime, their intensity and the induced electrical charge. Their influence on the gas and electrode temperatures is also evidenced through optical emission spectroscopy and infrared imaging. It is shown that, in our configuration, the conversion depends mostly on the charge delivered in the plasma and not on the effective plasma voltage when the applied power is modified. Similarly, at constant total current, a better conversion is observed at low frequencies, where a less filamentary discharge regime with a higher effective plasma voltage than that at a higher frequency is obtained. |

# 1. Introduction

Many industries still employ combustion processes from fossil fuels (natural gas, oil, coal, lignite, etc.), in particular the automotive and aerospace sectors, causing large emission of polluting and greenhouse gases [1]. As a result, carbon dioxide appears as one of the most problematic gases for the environment, mostly because it is responsible for global warming [2, 3].

To prevent the release of large quantities of $CO_2$ into the atmosphere [4], carbon capture and storage in geological reservoirs is one of the solutions which has been considered over the past decades. However, this approach presents several disadvantages including costs, environmental assessments and long-term risks. Moreover, it barely meets the principles of sustainable development since $CO_2$ is still considered a waste for long term storage. Nowadays, a more eco-friendly approach has emerged: $CO_2$ can be considered as a feedstock to convert into value-added products such as syngas, when mixing it with a H-source gas. In this respect, electrochemical and photochemical $CO_2$ reduction processes have already been successfully applied either to obtain CO, formic acid, formaldehyde, methanol and methane using electrical energy, or to obtain CO and formate-using light [5–7]. Also, splitting pure $CO_2$ into CO and $O_2$ is of interest since CO is more reactive than $CO_2$. It can be utilized in various catalytic processes such as the Monsanto process (combined with methanol to produce acetic acid) [8, 9] and the Fischer-Tropsch process (combined with H2 to produce liquid hydrocarbons) [10].

In recent years, cold plasmas have been pointed out in the research literature as a new approach for the valorization of $CO_2$ [11]. They operate far from thermodynamic equilibrium, i.e. the electrons are mostly heated by the applied electric power and can activate the gas by electron impact excitation, ionization and dissociation. New reactive species can, therefore, be created and converted more easily into new molecules, whilst the gas can remain at (or near) room temperature. Plasma can also allow the synthesis of specific organic molecules when combined with heterogeneous catalysis [12, 13]. For instance, $CO_2$–$CH_4$ treatments using Ni catalysts have already been used for the production of syngas and small amounts of organic molecules [14, 15]. Furthermore, plasma allows the possibility to store peak currents from sustainable energy into new fuels, since it can easily be switched on and off. Finally, the capture, transport and storage of $CO_2$ might become no more necessary, as dedicated plasma sources could allow their direct and online implementation on industrial smokestack outputs.

Various types of plasma sources have already been employed for $CO_2$ conversion, e.g. coronas [16, 17], gliding arcs [18–22], microwave plasmas [11, 23, 24] and dielectric barrier discharges (DBDs) [25–29]. In this article, the latter plasma source is investigated in a tubular configuration. The main advantage is the treatment of the entire gas flow, since all the gas passes through the active discharge region, in contrast to coronas and gliding arcs. Moreover, a DBD reactor has a simple design that can easily be up-scaled for industrial applications, as already demonstrated for the ozone synthesis [30–32]. We have already demonstrated that our flowing DBD setup presents a good compromise between a reasonable energy efficiency and a high $CO_2$ conversion for low specific energy input (SEI) [33]. Even if a DBD does not yet provide an economically feasible solution for $CO_2$ splitting [28, 34], it remains a promising approach when combined with catalysts in packed bed configuration. In such a case, it could allow for the selective production of value-added chemicals and also improve the energy efficiency of $CO_2$ splitting [27, 35–37]. Despite the economic limitations of a DBD for this type of application, it is important to better understand the electrical characteristics, and their effect on $CO_2$ dissociation.

The present study is performed for a pure $CO_2$ discharge in order to investigate the physical characteristics of the plasma. A complete study of the chemical mechanisms is beyond the scope of the present article but, as predicted by the simulations of Aerts et al, $CO_2$ splitting in a DBD plasma is mainly dictated by electron impact dissociation to form CO and O atoms [34]. If $CO_2$ is mixed with $CH_4$ to produce syngas (CO and $H_2$), the chemical pathways are more complicated [28, 38]. In some studies, $CO_2$ is also mixed with a rare gas such as argon or helium [33, 39–41]. In this case, no extra products are obtained, but the chemistry can still be affected, leading to a higher absolute $CO_2$ conversion with a lower effective conversion and energy efficiency [41]. Moreover, adding helium also affects the physical characteristics, because the discharge becomes more homogeneous, while adding argon still results in a filamentary discharge [33, 41].

In this experimental study, we investigate the influence of frequency and power on the filamentary behavior of a flowing DBD, operating at atmospheric pressure, to understand the role of the microdischarges in $CO_2$ conversion. For this purpose, we performed a detailed electrical characterization using a numerical method to extract data that are usually underestimated or poorly described in the research literature, i.e. the number and lifetime of microdischarges, the conduction current and electrical charge per half period or for a given residence time. Moroever, the temperature of the gas and the electrode are determined to obtain a better understanding of $CO_2$ splitting mechanisms.

## 2. Experimental methods

### 2.1. Reactor

A tubular DBD reactor dedicated to the treatment of $CO_2$ was designed, as shown in figure 1. It consists of an alumina tube, with inner and outer diameter of 26 and 30 mm, respectively, a concentric inner electrode powered by an AC high voltage, and an outer electrode, surrounding the alumina tube, which is grounded. The inner electrode is a copper rod with a diameter of 22 mm and a length of 120 mm, while the outer electrode is a stainless steel mesh that can easily be rolled around the alumina tube, which acts as dielectric barrier. The gap, i.e. the distance separating the inner electrode from the dielectric barrier, is 2 mm, while the length of the outer electrode is 100 mm, which defines the length of the plasma zone. This length is sufficient to ensure a long enough residence time. $CO_2$ was inserted in the reactor using a mass flow controller, keeping the flow rate fixed at 200 $mL_n.min^{-1}$. The gas enters via several inlets of 0.75 mm diameter, arranged into a circular pattern, then travels through the tubular reactor and finally flows out of the reactor via several outlets (with the same configuration as the inlet). The power applied to the high-voltage electrode is provided by an AFS generator G10S-V with a maximum power of 1000 W. This generator can be coupled to a transformer with the ability to tune the frequency in a range between 1 and 30 kHz. Figure 1 represents the entire experimental setup with all the diagnostic devices described in the following paragraphs.

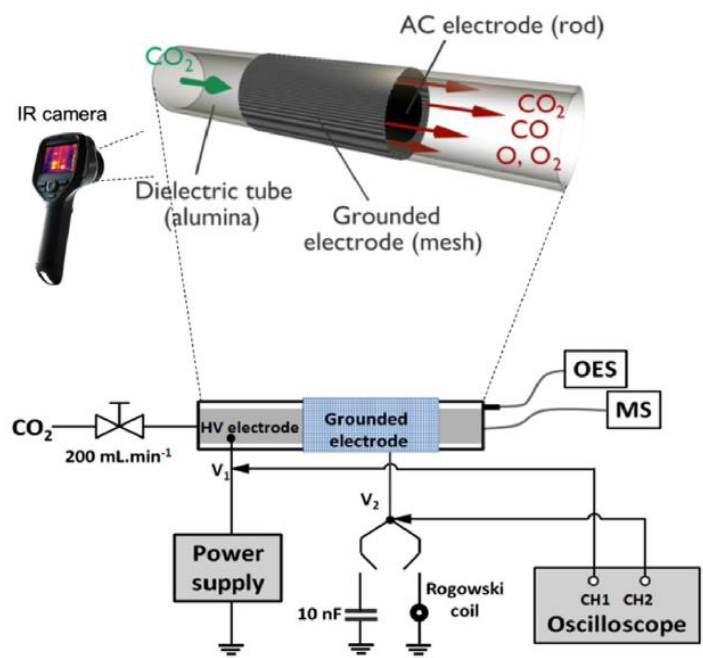

*Figure 1. Schematic diagram of the experimental set-up.*

### 2.2. Mass spectrometry (MS)

The gas flowing out of the DBD is characterized by a Hiden analytical QGA mass spectrometer (Warrington, UK), using a secondary electron multiplier (SEM) detector for an electron energy in the ionization chamber set to 70 eV. MASsoft7 software is used to monitor the partial pressure variations with specific m/z rations as a function of time.

## 2.3. Electrical measurements

All of the electrical measurements are performed using a digital oscilloscope (Tektronix DPO 3032). The applied voltage ($V_1$) was measured with a high voltage probe (Tektronix P6015A) while the DBD voltage is determined by subtracting $V_2$ from $V_1$, as represented in figure 1. An external capacitor (10 nF) was placed in series with the DBD to evaluate first its electrical charge $Q_P(t)$ and subsequently the power absorbed by the plasma via the Lissajous method [42]. The total current of the discharge ($I_{DBD}$) was recorded using a Pearson 2877 Rogowski coil placed in series with the DBD and presenting the ability to respond to fast changing currents due to its low inductance (rise time of 2 ns and drop rate of 0.2% $\mu s^{-1}$). The total current is the sum of the plasma current (i.e. conduction current) and dielectric current (i.e. capacitive current):

$$I_{DBD}(t) = I_{pl}(t) + I_{diel}(t)$$

At atmospheric pressure DBD in $CO_2$ usually operates in the filamentary mode [25, 34, 43, 44]. The microdischarges are characterized through their current peaks observed in the oscillograms (see figure 2) and considering the dielectric current $I_{diel}$ as the baseline for determining the microdischarge properties. Numerical methods from the Origin software analyzer toolbox were applied to determine the number of microdischarges ($N_{md}$), their peak intensity, electrical charge ($Q_{md}$) and lifetime ($L_{md}$). The intensity and lifetime of each microdischarge are evaluated considering the peak magnitude and FWHM, respectively. Its electric charge is determined by integrating the peak area from the baseline (see figure 2(c)). In the case of $N_{md}$, the 'filtering by height' method (including the local maximum option) is used. The accuracy of this numerical method is verified by comparing the provided $N_{md}$ value with the one measured from a visual counting. As the difference between the two approaches always remains lower than 5%, the numerical method is considered to be sufficiently accurate.

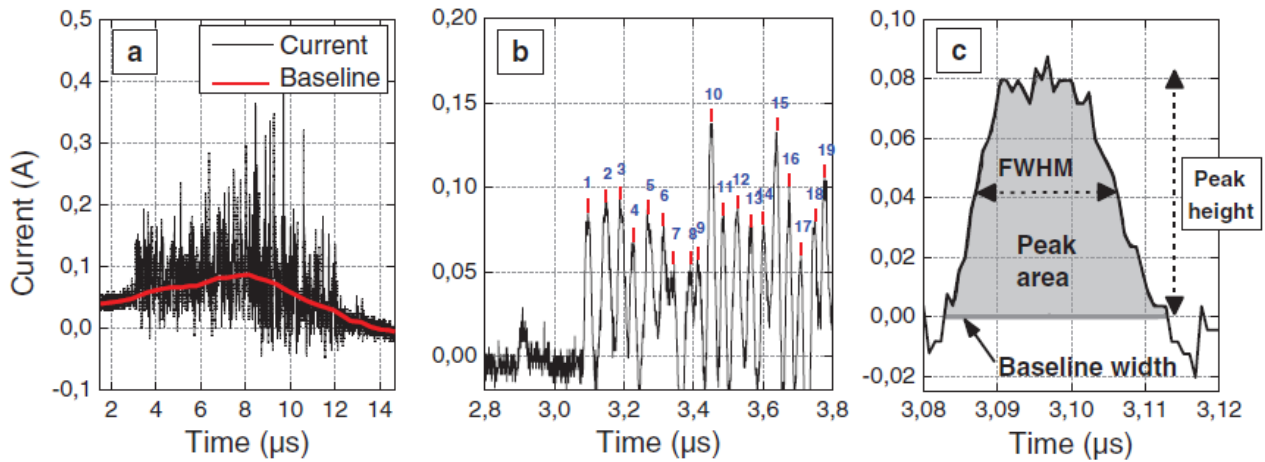

*Figure 2. (a) Oscillogram of the total current and determination of the baseline; (b) counting of the current peaks; (c) signal processing for the electrical characterization of the microdischarges.*

The main electrical parameters of the DBD were calculated and reported in table 1. Three different voltages were considered: the voltage over the entire DBD reactor ($V_{DBD}$), the voltage over the dielectric barrier ($V_{diel}$) calculated with $\xi$ as a dummy variable, and the effective voltage of the plasma ($V_{pl,eff}$). The electric charges of the DBD ($Q_{DBD}$), the dielectric barrier ($Q_{diel}$), the plasma ($Q_{pl}$) and the charge per micro-discharge ($Q_{md}$) are calculated from the current or voltage. Finally, the capacitances of the entire DBD ($C_{DBD}$), the dielectric barrier ($C_{diel}$) and the plasma ($C_{pl}$) are given, considering for $C_{diel}$ a dependence on the following parameters: the tubular barrier length (L), its external and internal radius ($R_{ext}$ and $R_{int}$, respectively) and the dielectric constant of alumina ($\varepsilon_r$). In the present work, the expression of the DBD capacitance is adopted from the homogeneous model of Godyak & Lieberman based on the equivalent circuit model of a capacitive discharge [45].

|  | Current | Voltage | Charge | Capacitance |
|---|---|---|---|---|
| Microdischarge | — | — | $Q_{md} = \int_{microdischarge} I_{pl}(t) \cdot dt$ | — |
| Plasma | $I_{pl}(t)$ | $V_{pl,eff}(t) = V_{DBD}(t) - V_{diel}(t)$ | $Q_{pl} = \int_\tau I_{pl}(t) \cdot dt$ | $C_{pl}$ |
| Dielectric barrier | $I_{diel}(t)$ | $V_{diel}(t) = \frac{1}{C_{diel}} \int_0^t I_{DBD}(\xi) \cdot d\xi + V_{diel}(0)$ | $Q_{diel} = C_{diel} \cdot V_{diel}$ | $C_{diel} = \frac{2\pi \varepsilon_0 \varepsilon_r \cdot L}{\mathrm{Ln}(R_{ext}/R_{int})}$ |
| DBD | $I_{DBD}(t) = I_{pl}(t) + I_{diel}(t)$ | $V_{DBD}(t) = V_1(t) - V_2(t)$ | $Q_{DBD} = \int_\tau I_{DBD}(t) \cdot dt$ | $\frac{1}{C_{DBD}} = \frac{1}{C_{pl}} + \frac{1}{C_{diel}}$ |

*Table 1. Main electrical parameters of the discharge.*

## 2.4. Optical emission spectroscopy (OES)

The OES measurements were performed with an Andor Shamrock-500i spectrometer including an Andor DU420A-OE CCD camera. Each spectrum was acquired for an exposure time of 5s and 7 accumulations. Silva et al have shown that in a $CO_2$ microwave plasma at low pressure, the rotational temperature of the CO band is similar to the gas temperature [46]. However, as the CO Angstrom band is not very suitable at atmospheric pressure for gas temperature measurements [47], we analyzed the first positive system (FPS) of $N_2$ by adding 10% of nitrogen to $CO_2$. Thereupon, the gas temperature has been measured via a line-ratio peak formula, specific to this system [48, 49]. The peaks at 775.3 and at 773.9 nm were considered.

## 2.5. Infrared imaging

2D temperature profiles of the grounded outer electrode were achieved with an infrared camera (FLIR E40) with a resolution of 160 × 120 pixels and a thermal sensitivity lower than 0.07 °C at 30 °C. FLIR ResearchIR software is used to control, record and analyze the temperature profiles in a range from −20 °C to +650 °C. The emissivity coefficient of the stainless steel electrode is about 0.45, and this value is introduced in the software. The temperature was calibrated to room temperature.

# 3. Results and discussion

## 3.1. Effect of the frequency

**3.1.1. $CO_2$ conversion and energy efficiency**

The effect of the frequency on the $CO_2$ conversion ($CO_2$) and on the $CO_2$ energy efficiency ($CO_2$) is investigated for an absorbed power fixed at 55 W (determined by the Lissajous method; applied power = 60 W). It was noticed that the absorbed power always remained constant for a given applied power of 60 W, regardless of the operating frequency used. $CO_2$ is calculated according to the following equation, where I stands for the mass spectrometer intensity assigned to $CO_2$:

$$\chi_{CO_2}(\%) = \frac{I_{CO_2\text{ plasma OFF}} - I_{CO_2\text{ plasma ON}}}{I_{CO_2\text{ plasma OFF}}} \times 100\% \quad (1)$$

The energy efficiency of the $CO_2$ conversion is calculated (in %) from $CO_2$, the enthalpy of the splitting reaction ($CO_2 \rightarrow CO + \frac{1}{2} O_2$), namely $\Delta H^0_{298K}$ = 279.8 kJ·mol$^{-1}$ = 2.9 eV·molecule$^{-1}$ and the SEI, according to the following equation:

$$\eta_{CO_2}(\%) = \chi_{CO_2}(\%) \cdot \frac{\Delta H^0_{298\,K}(\text{eV} \cdot \text{molecule}^{-1})}{\text{SEI (eV} \cdot \text{molecule}^{-1})} \quad (2)$$

The SEI corresponds to the energy density ($E_d$ in J·cm$^{-3}$), converted into eV·molecule$^{-1}$. The energy density depends on the power absorbed by the plasma and the gas flow rate.

$$E_d(\text{J} \cdot \text{cm}^{-3}) = \frac{\text{Plasma power (J} \cdot \text{s}^{-1})}{\text{Gas flow rate (cm}^3 \cdot \text{s}^{-1})} \quad (3)$$

$$\text{SEI (eV} \cdot \text{molecule}^{-1}) = \frac{E_d(\text{J} \cdot \text{cm}^{-3}) \times 6.24 \times 10^{18}(\text{eV} \cdot \text{J}^{-1}) \times 24\,500\,(\text{cm}^3 \cdot \text{mol}^{-1})}{6.022 \times 10^{23}(\text{molecule} \cdot \text{mol}^{-1})} \quad (4)$$

As shown in figure 3, an increase in the frequency from 16.2 to 28.6 kHz leads to a slight drop in the $CO_2$ conversion and energy efficiency. This is probably attributed to a drop in the density of electrons involved in the $CO_2$ splitting, although the total electron density in the whole discharge remains constant, as is shown below. The reason why there is a drop in conversion may be related to the fact that the shape of the electron energy distribution function (EEDF) is different when the plasma is in the filamentary regime (high frequency) or less filamentary regime-glow regime (low frequency) [33, 50]. Depending on the shape of the tail of EEDF, the number of electrons which can contribute to the breaking of the C = O bonds present in $CO_2$ might be higher for low frequency, i.e. electrons which are located on the right side of the EEDF over 5.52 eV (value of the C = O bond). Also, $CO_2$ > $CO_2$ in the whole frequency range because the SEI is in the order of ≈4.2 eV·molecule$^{-1}$, and thus always higher than $\Delta H_{298K}$.

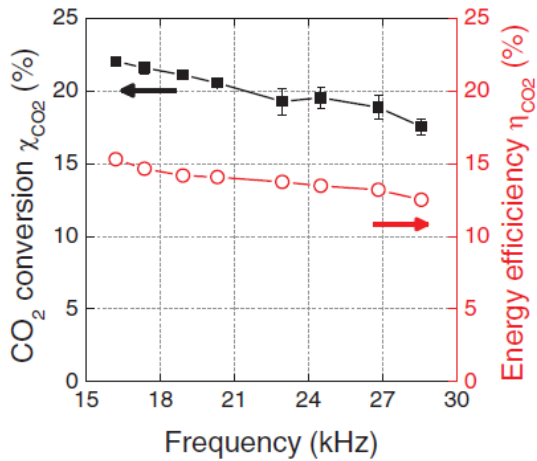
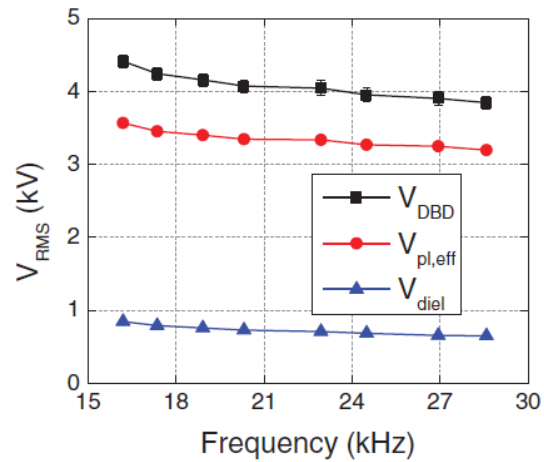

Figure 3. $CO_2$ conversion and energy efficiency as a function of the frequency, $P_{abs}$ = 55 W, $\Phi(CO_2)$ = 200 mL$_n$·min$^{-1}$.

Figure 4. RMS voltages as a function of frequency at $P_{abs}$ = 55 W, $\Phi(CO_2)$ = 200 mL$_n$·min$^{-1}$.

**3.1.2. Electrical characterization**

A complementary, electrical characterization was carried out to determine whether this trend in conversion and energy efficiency can be correlated with electrical measurements. As observed in figure 4, the DBD voltage ($V_{DBD}$) decreases as a function of the frequency (by almost 500 V in this frequency range).

For a DBD operating in its glow mode, the voltage is the sum of two components: plasma voltage ($V_{pl}$) and dielectric voltage ($V_{diel}$) [51–54]. As in our case the DBD operates in the filamentary mode, the presence of micro-discharges and remnant surface charges from previous micro-discharges leads to a potentially inhomogeneous electric field distribution. Therefore, the measured voltage has to be considered as an effective plasma voltage ($V_{pl,eff}$), i.e. a spatial average of the plasma voltage with potential local magnitude variations.

As the total electrical charge is conserved through the entire DBD source ($Q_{pl} = Q_{diel}$), the relation $C_{pl} \cdot V_{pl,eff} = C_{diel} \cdot V_{diel}$ leads to $V_{pl,eff}/V_{diel} = C_{diel}/C_{pl} = 11.2$. This ratio is rather high and this means that $V_{pl,eff}$ represents almost 82% of $V_{DBD}$, as represented by the red curve in figure 4. Similar values have also been reported in previous studies; for instance, in [44] $V_{pl,eff}$ was stated to represent even 98% of $V_{DBD}$. It is clear that $V_{DBD}$ and $V_{pl,eff}$ also decrease upon increasing frequency, and this can probably explain the drop in $CO_2$ conversion and energy efficiency, shown in figure 3 above, as a lower voltage might give rise to lower electron energies, and thus less efficient $CO_2$ conversion by electron impact excitation-dissociation. The dielectric voltage is obtained by subtracting $V_{pl,eff}$ from VDBD. It is almost constant in the entire frequency range with a value close to 450 V. To enhance the $CO_2$ conversion, one could increase $V_{pl,eff}$ (hence the electric field in the gap) by decreasing $V_{diel}$, e.g. by reducing the dielectric thickness.As the plasma current corresponds to a large number of individual peaks, and not to a continuous periodic function, studying its variation with respect to the frequency requires a dedicated analytical study. The number of peaks can be analyzed (i) per period or half period and (ii) for the entire $CO_2$ residence time or a given analysis time. In the present work, these variations are studied as follows:

(i) Per half period rather than per period since the current peak distributions can differ depending on their location on the positive or negative half periods.

(ii) For an analysis time fixed at 120 μs which is a more reasonable time window for this study than the residence time of $CO_2$. The latter one is estimated to 4.5 s considering the geometrical characteristics of the tubular discharge reactor and the $CO_2$ flow rate fixed at 200 $mL_n \cdot min^{-1}$.

The oscillograms of the total current ($I_{DBD} = I_{pl} + I_{diel}$) are plotted in figure 5 at 16.2 kHz, 22.9 kHz and 28.6 kHz. The comparison of these three oscillograms shows that increasing the frequency induces a decrease in the number of peaks per half period, which is not illogical, as the time of a half period drops upon higher frequency. Also, for a given frequency, the peak distribution in the positive half periods appears different from that in the negative half periods, which might be attributed to the asymmetry of the tubular discharge, where only one electrode is covered by a dielectric [55]. These oscillograms were analyzed using the aforementioned numerical method to determine the microdischarge characteristics with higher accuracy.

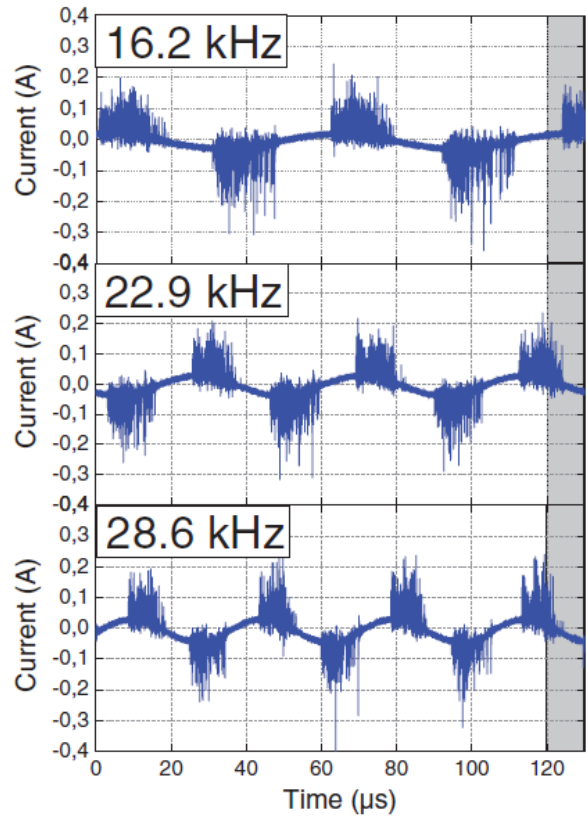

*Figure 5. Current profiles as a function of time for 3 different frequencies.*

The effect of the frequency on the number of microdischarges ($N_{md}$) is depicted in figure 6(a) per half period and for $\tau_{anal}$ = 120 µs. As mentioned above, a higher frequency results in a lower number of microdischarges per half period (whatever the sign of the half period) but it remains unchanged for $\tau_{anal}$ = 120 µs. Therefore, the total number of microdischarges cannot explain the decrease in the $CO_2$ conversion. The same applies to the average lifetime of the microdischarges illustrated in figure 6(b) as it remains close to 15 ns, regardless the operating frequency. As shown in figure 6(c), a rise in the frequency induces a slight drop in the plasma charge per half period (with $Q_{pl}^+$ = $Q_{pl}^-$). As the effect of the frequency on $CO_2$ has to be explained based on the residence time, it is more relevant to consider $Q_{DBD}^{120}$, as plotted in figure 6(d). In this figure, the electrical charge of the DBD increases with frequency due to its dielectric component ($Q_{diel}$) and not due to its plasma component ($Q_{pl}$). In other words, the frequency does not seem to change the electrical properties of the plasma but the electric field distribution through the dielectric barrier.

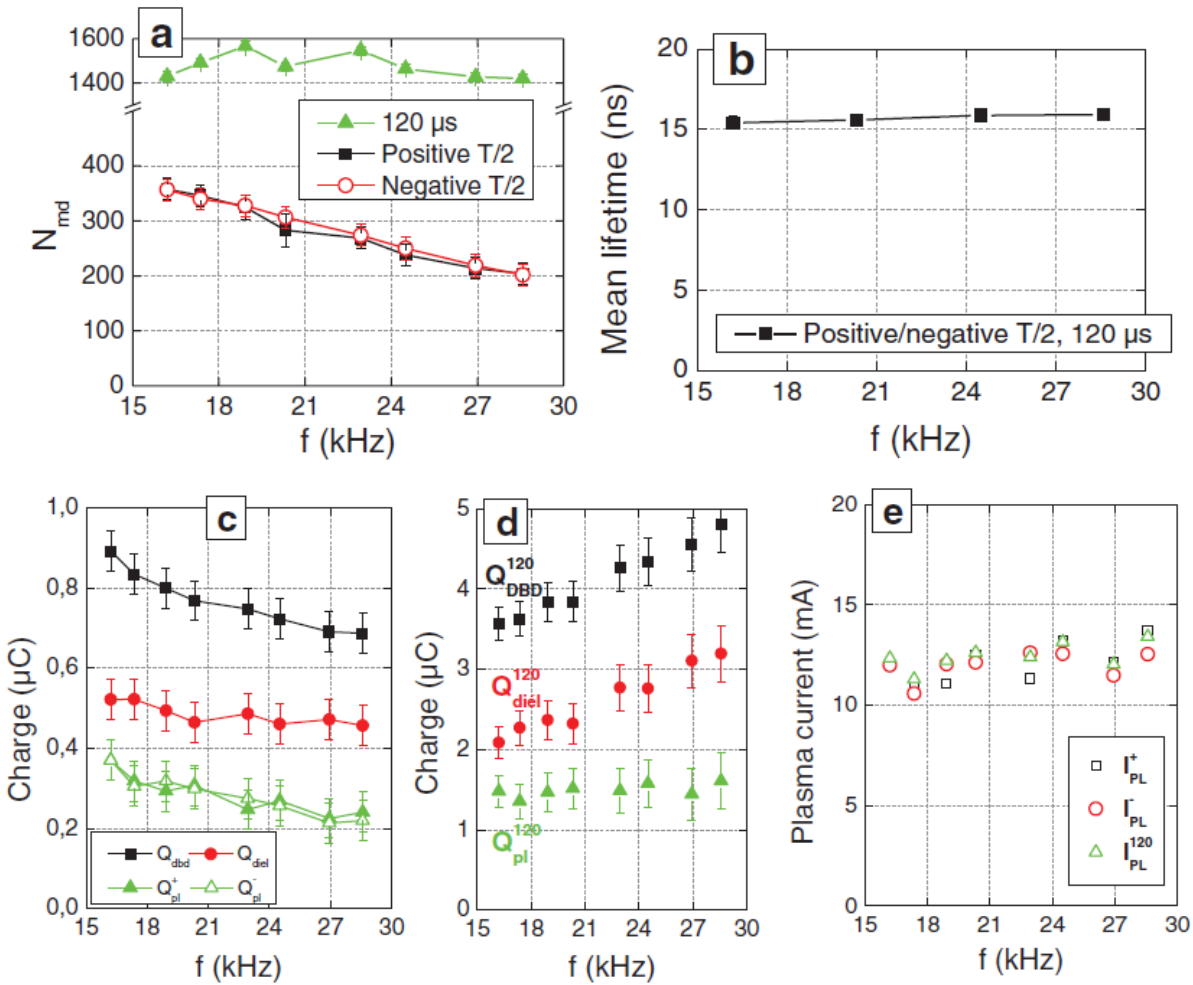

*Figure 6.* (a) Number, (b) lifetime of the microdischarges per half period (T/2) and for $\tau_{anal}$ = 120 µs; (c) charge per T/2, (d) charge for $\tau_{anal}$ =120 µs and (e) conduction current per T/2 and for $\tau_{anal}$ =120 µs.

Also, $Q_{pl}^+$, $Q_{pl}^-$ and $Q_{pl}^{120}$ can be easily converted into conduction currents considering either the half period or the analysis time, as expressed in the following formulas:

$$I_{pl}^+(f) = \frac{Q_{pl}^+(f)}{T/2} = 2f \cdot Q_{pl}^+(f)$$

$$I_{pl}^-(f) = \frac{Q_{pl}^-(f)}{T/2} = 2f \cdot Q_{pl}^-(f)$$

$$I_{pl}^{120}(f) = \frac{Q_{pl}^{120}(f)}{\tau_{anal}}$$

Plotting these currents in figure 6(e) indicates that their variations versus the frequency are not that very significant: only a slight increase of 20% (from 11 mA to 14 mA) is observed for a half period. The discharge current also increases slightly for a given residence time. This is logical, as the voltage slightly drops upon rising frequency, so the current should slightly increase to keep the same power. Hence, it is clear that the microdischarge characteristics are not correlated to $CO_2$ conversion.

An alternative and convenient representation to evaluate how the frequency could change the microdischarge characteristics is to plot the distribution of their lifetimes versus their number ($N_S$) for several frequencies, as presented in figure 7. Most of the microdischarges ($N_S \approx 1200$) exhibit a lifetime comprised between 7 and 25 ns. A small fraction of microdischarges exhibits a lifetime under 7 ns and above 25 ns. The drop in $CO_2$ conversion upon rising frequency may not be induced by the microdischarges. Only the drop of the discharge voltage seems to be responsible for the drop in $CO_2$ conversion.

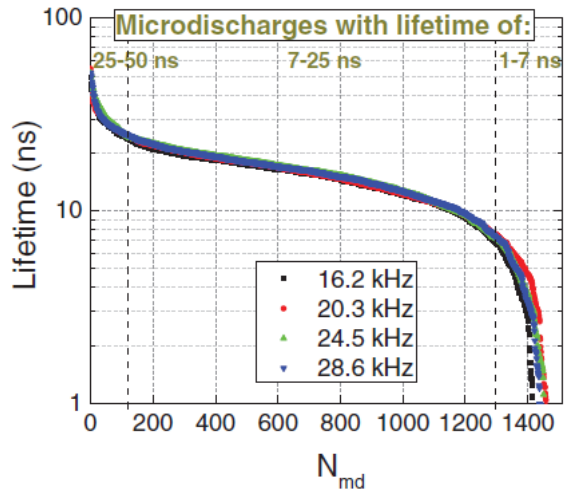

*Figure 7. Distribution of the microdischarges lifetime for $\tau_{anal} = 120$ μs and $P_{abs} = 55$ W*

### 3.1.3. Gas temperature and electrode temperature

A slight increase in the conduction current as well as in the microdischarge lifetime may have impact on the power dissipated by the discharge by the Joule effect since it is proportional to the square of $I_{DBD}$. This assumption has been verified by investigating the influence of the frequency on the gas temperature (obtained from OES measurements) as well as on the electrode temperature (obtained from IR imaging). Figure 8(a) shows the 2D temperature profiles of the outer grounded electrode, obtained by placing the IR camera perpendicular to the gas flow. The electrode temperature appears rather uniform in the whole discharge region, regardless of the frequency. As illustrated in figure 8(b), increasing the frequency from 16.2 kHz to 28.6 kHz induces an increase in the gas temperature (from 523 to 560 K) and a slight increase in the average temperature of the ground electrode (from 375 K to 405 K). The difference $T_{gas}$–$T_{electrode}$ can be considered as a representation of Joule losses. A rise in the frequency induces an increase in the gas temperature that can be linked to a decrease of the effective plasma voltage (see figure 4). Indeed, a low frequency plasma gives place to a higher effective plasma voltage, i.e. to a higher electric field where electrons can be more readily accelerated. The increase in gas temperature upon rising frequency may be correlated with the lower $CO_2$ conversion at high frequencies since more energy may be spent for heating and less to decompose $CO_2$. The gas temperature values which are found with the OES spectra are in good agreement, and in the same order of magnitude, with temperature values determined by other methods for a $CO_2$ DBD in the same kind of conditions [56].

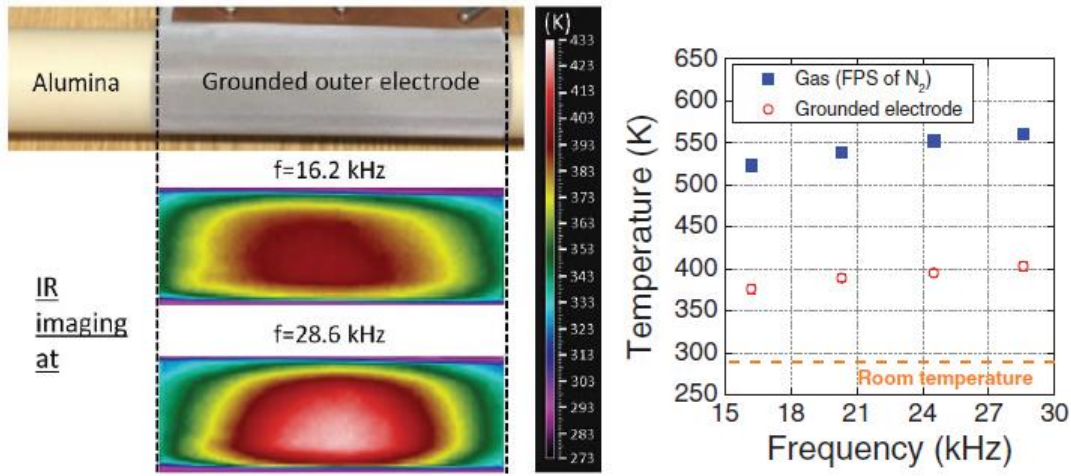

Figure 8. (a) 2D temperature profiles of the grounded electrode; (b) Gas and grounded electrode temperature as a function of frequency. $P_{abs}$ = 55 W, process time = 5 min.

## 3.2. Effect of the power

**3.2.1. $CO_2$ conversion and energy efficiency**

Figure 9 shows that the $CO_2$ conversion rises linearly with power, reaching a value as high as 28% at 100 W. In the same power range, a slight decrease in the energy efficiency is measured from 12.5% to 11%. These trends are logical, because more energy is put into the system, leading to a higher electron density [57, 58], and to more $CO_2$ conversion. At the same time, the rise in $CO_2$ conversion is less pronounced than the rise in SEI, thus the energy efficiency drops upon increasing power, as predicted by equation (2). An electrical characterization was carried out to determine whether this conversion trend is related to specific properties of the filaments.

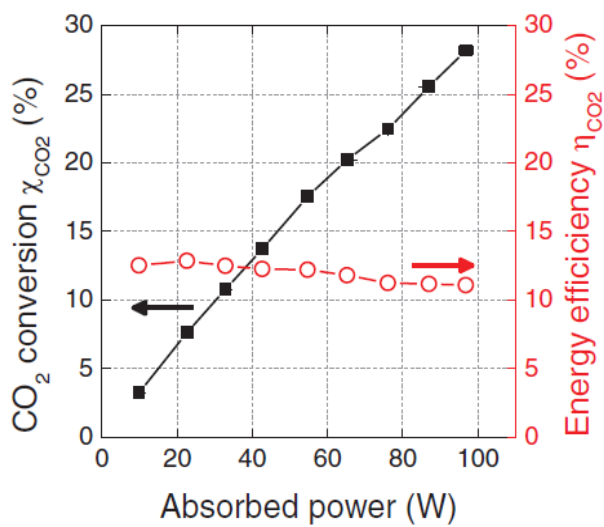

Figure 9. $CO_2$ conversion and energy efficiency as a function of the absorbed power, for f = 28.6 kHz, and $\Phi(CO_2)$ = 200 $mL_n·min^{-1}$.

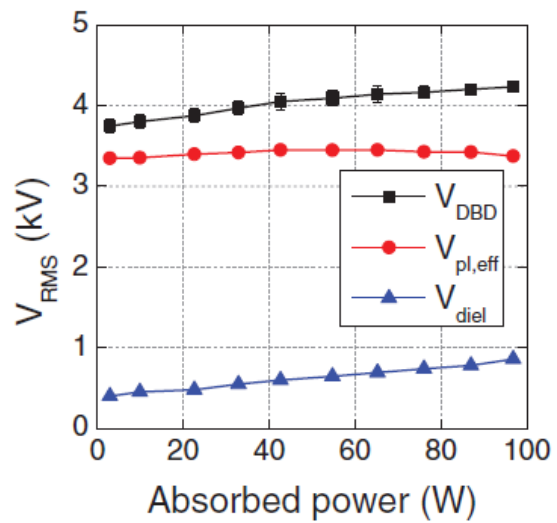

Figure 10. RMS voltages as a function of the absorbed power, for f = 28.6 kHz and $\Phi(CO_2)$ = 200 $mL_n·min^{-1}$.

**3.2.2. Electrical characterization**

Firstly, the influence of the absorbed power on the DBD voltage was studied, as plotted in figure 10, which shows a clear increase from 3.8 kV to 4.2 kV, as expected. $V_{pl,eff}$ represents 89% of the DBD voltage at 3 W and it drops to 80% at 97 W. The effective plasma voltage is thus not really correlated with the increase in $CO_2$ but the influence of the filaments has to be investigated as well for a complete interpretation of the behavior.

As the oscillograms present a very symmetric signal for the positive and negative sign of T/2, we only focused our study on the filaments for a certain analysis time. Figure 11(a) depicts the number of microdischarges within a fixed $\tau_{anal}$, as well as their mean lifetime, as a function of the absorbed power. Increasing the power leads to a rise in the number of peaks as well as their mean lifetime. The power effect is, therefore, very different from the frequency effect where no influence was observed both on $N_{md}$ and the microdischarge lifetime (see figures 6(a) and (b)). Also, figure 11(b) illustrates the electrical charge of the DBD versus the absorbed power: a clear increase was observed from 20 W to 100 W. This is attributable to the increase in the plasma charge (which corresponds therefore to an increase of the discharge current), even if the dielectric component remains always higher than the plasma component in the entire power range. The latter one, however, remained rather constant because the operating frequency is fixed. Thus, it is clear that the rise in number of microdischarges, in the microdischarge lifetime, and in the total electric charge of the DBD, are correlated with the rise in $CO_2$ conversion with higher power.

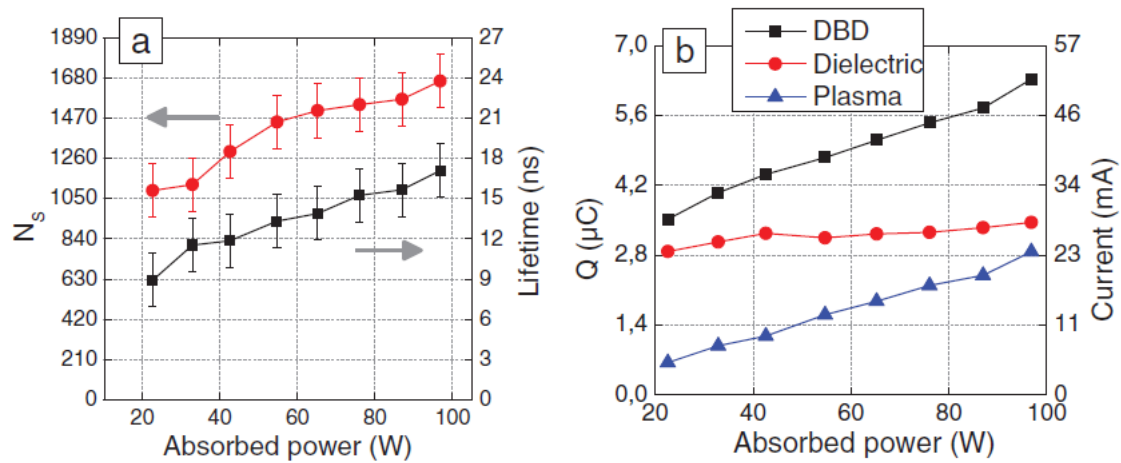

*Figure 11. Characteristics of the microdischarges as a function of the absorbed power:*
*(a) number of current peaks and mean lifetime (b) charge and current for $\tau_{anal}$ = 120 μs.*

The distribution of the microdischarge lifetimes versus their number ($N_{md}$) is represented in figure 12. It appears that decreasing the power leads mostly to the production of short lifetime microdischarges (cf. line a). Indeed, the percentage of microdischarges which has less than 7 ns of lifetime is 31, 23 and 8% for a plasma power of 23, 65 and 97 W, respectively. However, increasing the power leads to a greater extent to the production of long-lifetime microdischarges (cf. line b). Indeed, the percentage of microdischarges which has more than 25 ns of lifetime is 3, 5 and 12% for a plasma power of 23, 65 and 97 W, respectively. This effect is not observed when tuning the frequency.

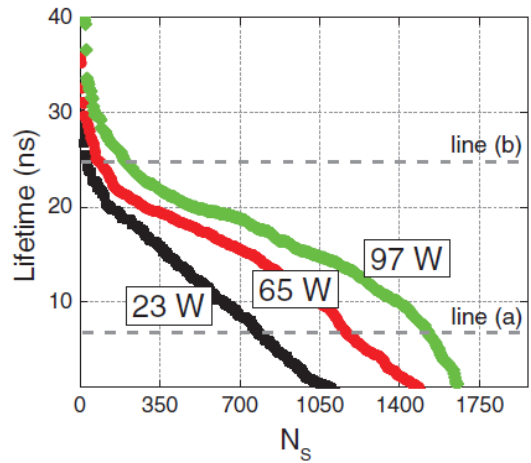

*Figure 12. Distribution of the microdischarges lifetime for f = 28.6 kHz, $\tau_{anal}$ = 120 µs.*

### 3.2.3. Gas temperature and electrode temperature

Furthermore, we also checked whether the rising $CO_2$ conversion is correlated with the gas temperature and the temperature of the grounded electrode. In figure 13(a), the 2D profiles of the temperature at the outer grounded electrode are plotted for 20 W, 50 W, 80 W and 100 W. The electrode temperature appears not entirely uniform in the whole discharge region, and it is clearly correlated with power. As illustrated in figure 13(b), the temperature of the grounded electrode increases linearly with power and reaches a mean value as high as 430 K at 100 W after 5 min of process time. On the other hand, the gas temperature increases by almost 40 K as a function of the power as can be observed in figure 13(b). Increasing the plasma power generally enhances the electron impact processes and, thus, could lead to gas heating. When the plasma is ignited, the gas heating is responsible for an increase of the wall temperature. The Joule losses are quite significant since both the alumina dielectric and the mesh stainless steel electrode have quite high thermal conductivities.

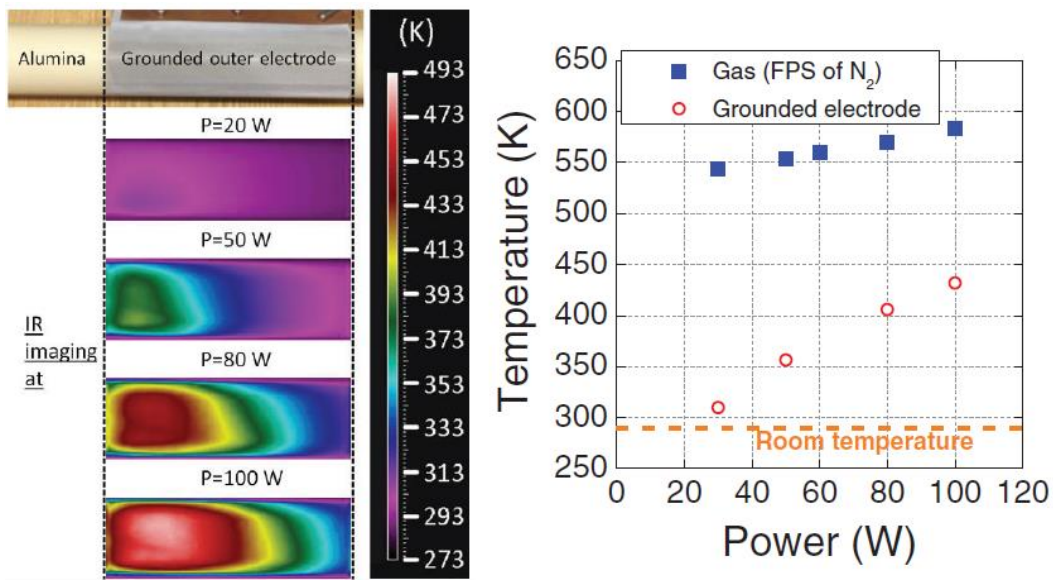

*Figure 13. (a) 2D profiles of the grounded electrode temperature for various powers;*
*(b) Gas temperature and grounded electrode temperature as a function of power (f = 28.6 kHz, process time = 5 min).*

# 4. Conclusion

The splitting of $CO_2$ into CO and $O_2$ was studied at atmospheric pressure in a tubular DBD operating in pure $CO_2$. The influence of the operating frequency and applied plasma power on the discharge behavior and on $CO_2$ conversion and energy efficiency was discussed.

Table 2 summarizes the results of all the experiments to correlate the $CO_2$ conversion with the process parameters by increasing the frequency and the power. We can observe how the frequency and power affect the $CO_2$ conversion and energy efficiency, as well as other plasma parameters, for $\tau_{anal}$ = 120 μs. It is clear that these two parameters, which can easily be controllable on the high voltage AC generator, have an important impact on the $CO_2$ splitting process. It was observed that both parameters (especially the power) strongly affect the charge delivered in the plasma, the number of microdischarge which occurs for a certain time, the gas temperature, the effective plasma voltage, the total number of electron and also the number of electron which has enough energy to split $CO_2$, depending on the shape of the EEDF and, thus, defining the energy efficiency of the splitting process.

|  |  |  | Increase in the frequency (from 16.2 to 28.6 kHz) | Increase in the power (from 3 to 97 W) |
|---|---|---|---|---|
| $CO_2$ conversion ($\chi_{CO_2}$) |  |  | ↘ (22–17.5%) | ↘ (3.2–28.2 %) |
| $CO_2$ energy efficiency ($\eta_{CO_2}$) |  |  | ↘ (15.3–12.5 %) | ↘ (12.5–11.1 %) |
| Voltage | Plasma |  | ↘ (3.57–3.20 kV) | ≈ (3.34–3.37 kV) |
|  | Dielectric |  | ↘ (0.85–0.65 kV) | ↗ (0.40–0.86 kV) |
| Current peaks | Number |  | = | ↗ |
|  | Lifetime |  | ≈ (15.4–15.9 ns) | ↗ (8.9–17.2 ns) |
|  | Intensity |  | ≈ (12.3–13.4 mA) | ↗ (5.5–24.0 mA) |
|  | Distribution |  | Unchanged | Changed |
| Charge | Plasma ($Q_{pl}$) |  | ≈ | ↗ |
|  | DBD ($Q_{DBD}$) |  | ↗ | ↗ |
| Temperature | Gas ($T_g$) |  | ↗ (523–560 K) | ↗ (543–583 K) |
|  | Electrode |  | ↗ (376–403 K) | ↗ (309–432 K) |

*Table 2. Summary of interpretation for the $CO_2$ splitting.*

Increasing the frequency may induce an increase in $T_e$, accounting for the measured increase in $T_{gas}$. Moreover, the gas heating also leads to a slight increase in the outer electrode temperature. As the microdischarge distributions do not depend on the frequency, and the same applies to the microdischarge average lifetime, the number and electrical charge of the microdischarges, we consider that the microdischarges do not affect the $CO_2$ conversion in the frequency study. The most relevant parameter to explain the decrease in $CO_2$ with rising frequency is the drop of the effective plasma voltage. The increase in the gas temperature with the power may itself result from a rise in the electron impact processes. This could explain the increase of the $CO_2$ conversion in close correlation with the electrical characterizations.

# 5. Acknowledgments

The authors acknowledge financial support from the IAPVII/12, P7/34 (Inter-university Attraction Pole) program 'PSI-Physical Chemistry of Plasma-Surface Interactions', financially supported by the Belgian Federal Office for Science Policy (BELSPO). A Ozkan would like to thank the financial support given by 'Fonds David et Alice Van Buuren'.

# 6. References

[1] Gurney K R, Mendoza D L, Zhou Y, Fischer M L, Miller C C, Geethakumar S and de la Rue du Can S 2009 Environ. Sci. Technol. 43 5535–41

[2] Meinshausen M, Smith S J, Calvin K, Daniel J S, Kainuma M, Lamarque J, Matsumoto K, Montzka S, Raper S and Riahi K 2011 Clim. Change 109 213–41

[3] Zecchina A 2014 Rend. Lincei 25 113–7

[4] Feron P and Hendriks C 2005 Oil Gas Sci. Technol. 60 451–9

[5] Aresta M 2010 Carbon Dioxide as Chemical Feedstock (New York: Wiley)

[6] Christophe J, Doneux T and Buess-Herman C 2012 Electrocatalysis 3 139–46

[7] Sakakura T, Choi J-C and Yasuda H 2007 Chem. Rev. 107 2365–87

[8] Jones J H 2000 Platinum Met. Rev. 44 94–105

[9] Sunley G J and Watson D J 2000 Catal. Today 58 293–307

[10] Dry M E 2002 Catal. Today 71 227–41

[11] Spencer L F and Gallimore A D 2011 Plasma Chem. Plasma Process. 31 79–89

[12] Fan M S, Abdullah A Z and Bhatia S 2009 ChemCatChem 1 192–208

[13] Scapinello M, Martini L M and Tosi P 2014 Plasma Process. Polym. 11 624–8

[14] Tu X, Gallon H J, Twigg M V, Gorry P A and Whitehead J C 2011 J. Phys. D: Appl. Phys. 44 274007

[15] Tu X and Whitehead J 2012 Appl. Catal. B: Environ. 125 439–48

[16] Ghorbanzadeh A, Lotfalipour R and Rezaei S 2009 Int. J. Hydrog. Energy 34 293–8

[17] Li D, Li X, Bai M, Tao X, Shang S, Dai X and Yin Y 2009 Int. J. Hydrog. Energy 34 308–13

[18] Bo Z, Yan J, Li X, Chi Y and Cen K 2008 Int. J. Hydrog. Energy 33 5545–53

[19] Indarto A, Yang D R, Choi J-W, Lee H and Song H K 2007 J. Hazardous Mater. 146 309–15

[20] Nunnally T, Gutsol K, Rabinovich A, Fridman A, Gutsol A and Kemoun A 2011 J. Phys. D: Appl. Phys. 44 274009

[21] Pornmai K, Jindanin A, Sekiguchi H and Chavadej S 2012 Plasma Chem. Plasma Process. 32 723–42

[22] Tu X and Whitehead J C 2014 Int. J. Hydrog. Energy 39 9658–69

[23] Spencer L and Gallimore A 2013 Plasma Sources Sci. Technol. 22 015019

[24] Vesel A, Mozetic M, Drenik A and Balat-Pichelin M 2011 Chem. Phys. 382 127–31

[25] Brehmer F, Welzel S, van de Sanden M and Engeln R 2014 J. Appl. Phys. 116 123303

[26] Duan X, Li Y, Ge W and Wang B 2015 Greenhouse Gases: Sci. Technol. 5 131–40

[27] Mei D, Zhu X, He Y-L, Yan J D and Tu X 2015 Plasma Sources Sci. Technol. 24 015011

[28] Snoeckx R, Zeng Y, Tu X and Bogaerts A 2015 RSC Adv. 5 29799–808

[29] Yu Q, Kong M, Liu T, Fei J and Zheng X 2012 Plasma Chem. Plasma Process. 32 153–63

[30] Eliasson B, Hirth M and Kogelschatz U 1987 J. Phys. D: Appl. Phys. 20 1421

[31] Kogelschatz U 2003 Plasma Chem. Plasma Process. 23 1–46

[32] Kogelschatz U, Eliasson B and Egli W 1999 Pure Appl. Chem. 71 1819–28

[33] Ozkan A, Dufour T, Arnoult G, De Keyzer P, Bogaerts A and Reniers F 2015 J. CO2 Utilization 9 74–81

[34] Aerts R, Somers W and Bogaerts A 2015 ChemSusChem 8 702–16

[35] Tu X, Gallon H J and Whitehead J C 2011 IEEE Trans. Plasma Sci. 39 2172–3

[36] Van Laer K and Bogaerts A 2015 Energy Technol. 3 1038–44

[37] Zhang Y, Wang H-y, Jiang W and Bogaerts A 2015 New J. Phys. 17 083056

[38] Snoeckx R, Aerts R, Tu X and Bogaerts A 2013 J. Phys. Chem. C 117 4957–70

[39] Pinhao N, Janeco A and Branco J 2011 Plasma Chem. Plasma Process. 31 427–39

[40] Goujard V, Tatibouët J-M and Batiot-Dupeyrat C 2011 Plasma Chem. Plasma Process. 31 315–25

[41] Ramakers M, Michielsen I, Aerts R, Meynen V and Bogaerts A 2015 Plasma Process. Polym. 12 755–63

[42] Manley T 1943 Trans. Electrochem. Soc. 84 83–96

[43] Aerts R, Martens T and Bogaerts A 2012 J. Phys. Chem. C 116 23257–73

[44] Valdivia-Barrientos R, Pacheco-Sotelo J, Pacheco-Pacheco M, Benitez-Read J and López-Callejas R 2006 Plasma Sources Sci. Technol. 15 237

[45] Lieberman M A and Lichtenberg A J 1994 Principles of Plasma Discharges and Materials Processing (New York: Wiley)

[46] Silva T, Britun N, Godfroid T and Snyders R 2014 Plasma Sources Sci. Technol. 23 025009

[47] Bruggeman P, Sadeghi N, Schram D and Linss V 2014 Plasma Sources Sci. Technol. 23 023001

[48] Britun N, Godfroid T, Konstantinidis S and Snyders R 2011 Appl. Phys. Lett. 98 141502

[49] Britun N, Godfroid T and Snyders R 2012 Plasma Sources Sci. Technol. 21 035007

[50] Godyak V A 2006 IEEE Trans. Plasma Sci. 34 755

[51] Liu S and Neiger M 2001 J. Phys. D: Appl. Phys. 34 1632

[52] Liu S and Neiger M 2003 J. Phys. D: Appl. Phys. 36 3144

[53] Massines F, Gherardi N, Naude N and Segur P 2005 Plasma Phys. Control. Fusion 47 B577

[54] Tao S, Kaihua L, Cheng Z, Ping Y, Shichang Z and Ruzheng P 2008 J. Phys. D: Appl. Phys. 41 215203

[55] Belov I, Paulussen S and Bogaerts A Plasma Sources Sci. Technol. 25 015023

[56] Brehmer F, Welzel S, Klarenaar B, van der Meiden H, van de Sanden M and Engeln R 2015 J. Phys. D: Appl. Phys. 48 155201

[57] Shrestha R, Tyata R and Subedi D 2013 Himalayan Phys. 4 10–13

[58] Wu A J, Zhang H, Li X D, Lu S Y, Du C M and Yan J H 2015 IEEE Trans. Plasma Sci. 43 836–45